\documentclass[aps,prl,superscriptaddress,twocolumn]{revtex4-2}

\usepackage{comment}
\usepackage[utf8]{inputenc}
\usepackage{amssymb}
\usepackage{amsmath}
\usepackage{amsfonts}
\usepackage{bm}
\usepackage{mathtools}
\usepackage{bbold}
\usepackage{upgreek}
\usepackage{xfrac}
\usepackage{soul}
\usepackage{bm}
\usepackage{xcolor}
\usepackage{derivative}
\usepackage{physics}

\usepackage[colorlinks=true,citecolor=blue]{hyperref}
\hypersetup{colorlinks=true,citecolor=blue,linkcolor=blue,urlcolor=blue}
\usepackage{url}

\usepackage{txfonts}
\DeclareSymbolFont{matha}{OML}{txmi}{m}{it}% txfonts
\DeclareMathSymbol{\varv}{\mathord}{matha}{118}
\usepackage{bbm}

 \makeatletter
\def\@fnsymbol#1{\ensuremath{\ifcase#1\or \dagger\or \ddagger\or
   \mathsection\or \mathparagraph\or \|\or **\or \dagger\dagger
   \or \ddagger\ddagger \else\@ctrerr\fi}}
    \makeatother

\DeclareFontEncoding{LS1}{}{}
\DeclareFontSubstitution{LS1}{stix}{m}{n}

\begin{document}

\title{Multidimensional Quantum Generative Modeling by Quantum Hartley Transform}

\author{Hsin-Yu Wu}
\email{h.wu@exeter.ac.uk}
\affiliation{Department of Physics and Astronomy, University of Exeter, Stocker Road, Exeter EX4 4QL, United Kingdom}
\affiliation{PASQAL, 7 Rue Léonard de Vinci, 91300 Massy, France}

\author{Vincent E. Elfving}
\affiliation{PASQAL, 7 Rue Léonard de Vinci, 91300 Massy, France}

\author{Oleksandr Kyriienko}
\affiliation{Department of Physics and Astronomy, University of Exeter, Stocker Road, Exeter EX4 4QL, United Kingdom}
\affiliation{PASQAL, 7 Rue Léonard de Vinci, 91300 Massy, France}

\date{\today}

\begin{abstract}
We develop an approach for building quantum models based on the exponentially growing orthonormal basis of Hartley kernel functions. First, we design a differentiable Hartley feature map parametrized by real-valued argument that enables quantum models suitable for solving stochastic differential equations and regression problems. Unlike the naturally complex Fourier encoding, the proposed Hartley feature map circuit leads to quantum states with real-valued amplitudes, introducing an inductive bias and natural regularization. Next, we propose a quantum Hartley transform circuit as a map between computational and Hartley basis. We apply the developed paradigm to generative modeling from solutions of stochastic differential equations, and utilize the quantum Hartley transform for fine sampling from parameterized distributions through an extended register. Finally, we present tools for implementing multivariate quantum generative modeling for both correlated and uncorrelated distributions. As a result, the developed quantum Hartley models offer a distinct quantum approach to generative AI at increasing scale. 

%providing an efficient readout in the bitstring basis
\end{abstract}

\maketitle

\section*{Introduction}
Quantum computing offers speedup in solving certain problems, with a promise of outperforming classical solvers thanks to natively quantum effects of entanglement and superposition~\cite{NielsenChuang}. Quantum algorithms that take advantage of these properties cover various application areas, ranging from cryptography~\cite{shor1994algorithms} and quantum simulation~\cite{Lloyd1996,Bulata2009}, to linear equation solvers~\cite{harrow2009,ambainis2010,Berry_2014} and optimization~\cite{farhi2014quantum,lloyd2018quantum,Hadfield2019}. Most of the protocols mentioned above rely on subroutines that involve the quantum phase estimation and the quantum Fourier transform (QFT)~\cite{kitaev1995quantum,Cleve_1998,Hales2000,coppersmith2002approximate}. The QFT is the implementation of the classical discrete Fourier transform (DFT) on a quantum circuit (QC) and it can be derived in an equivalent way from a fast Fourier transform based on the Danielson-Lanczos lemma~\cite{FFT1965,Press2007,Agaian_2002}. Using the phase kickback trick, one can show that both resulting QFT circuits are identical. As the vast majority of applications is concerned with the processing of real-valued datasets, the Hermitian-symmetric QFT manifests itself that there is redundancy in the spectral expansion of a real-valued signal. In addition, one complex multiplication requires the application of four real multiplications and three additions/subtractions, while one real multiplication just needs two multiplications and one addition/subtraction. Consequently, the use of the Fourier-related algorithms capable of naturally performing real-valued unitary transformations between the real and reciprocal spaces is highly desirable. Namely, it adds an inductive bias, thus reducing resources required for building the models, and helping to prepare quantum probability distributions. %This is relevant for near-term models ~\cite{Preskill2018}.
%% since it saves time by omitting the computation of imaginary parts, requires only half the memory storage and executes fewer operations that may lead to fewer rounding and truncation errors, especially for noisy intermediate-scale quantum (NISQ) devices~\cite{Preskill2018}.

Quantum machine learning (QML) is a burgeoning interdisciplinary field that integrates quantum computing with machine learning. Usually it refers to the use of variational quantum algorithms to undertake classical learning tasks and solve practically relevant problems ~\cite{peruzzo2014variational,schuld2015,Kandala_2017,Biamonte2017,Benedetti2019,Lubasch2020,Biamonte2021,Cerezo_2021,BravoPrieto2023}. Motivated by the great success of classical generative models in machine learning~\cite{sohldickstein2015deep,goodfellow2020,ho2020ddpm,song2021scorebased}, various protocols for quantum generative modeling (QGM) have recently been developed to exploit parameterized QCs to represent observable-dependent (implicit) or data-dependent (explicit) probability distributions~\cite{Liu2018,PerdomoOrtiz2018,Benedetti2019npj,Coyle2020,Havl_ek_2019,Schuld2019QML,lloyd2020quantum,Paine2021,Kyriienko2022,kasture2022protocols,rudolph2023trainability}. A prominent example of the implicit quantum models is the so-called quantum circuit Born machine (QCBM), which is constructed based on Born's rule~\cite{Liu2018}. As soon as QCBM is successfully trained using data represented by binary strings (thus building an implicit model \cite{rudolph2023trainability}), one can proceed directly to sampling from the same circuit. By contrast, the explicit quantum models rely on a feature map encoding of continuous or discrete distributions of data~\cite{kyriienko2021solving,Kyriienko2022,kasture2022protocols}. A typical workflow for the quantum explicit models starts with encoding classical input data of the form $\bm{x} = \{x_0, \cdots, x_{2^N-1}\}$ to quantum states via a $N$-qubit quantum feature map circuit that acts on a zero product state $|\mathrm{\o}\rangle \equiv |0\rangle^{\otimes N}$. Specifically, a feature map is a unitary operator that takes an input argument $x$ and maps it to a distinct quantum state living in a $2^N$ Hilbert space, $x \mapsto |\varphi(x)\rangle = \hat{\mathcal{U}}_{\varphi}(x) |\mathrm{\o}\rangle$. One feature map that associates an input feature with a quantum state in the phase space is the Fourier (phase) feature map \cite{Kyriienko2022}. It is formed by an initial layer of Hadamard gates on each qubit followed by a layer of scaled phase shift gates, $\mathrm{P}_l(x)=\mathrm{diag}\{1,\exp(i 2 \pi x /2^l)\}$ applied to the qubit index $l \in [1, \cdots,N]$. The set of the output states $\{|\varphi(x_j)\rangle \}_{j=0}^{2^N-1}$, evaluated at $\{x_j\}_{j=0}^{2^N-1}$ representing consecutive integers in the range of $[0, 2^N-1]$ corresponding to length-$N$ binary strings, forms a complete orthonormal Fourier basis. These Fourier basis states can be mapped to the set of computational basis through an inverse quantum Fourier transform (QFT) circuit, $\{|x_j\rangle \}_{j=0}^{2^N-1} = \hat{\mathcal{U}}_{\mathrm{QFT}}^{\dagger} \{|\varphi(x_j)\rangle \}_{j=0}^{2^N-1} $. Unlike other quantum encoding techniques, such as amplitude and basis embeddings~\cite{MariaSchuldbook,lloyd2020embeddings}, $\hat{\mathcal{U}}_{\varphi}(x)$ can be differentiated with respect to any continuous variable $x \in \mathbb{R}_{2^N-1}$, allowing solving differential equations modeled in the explicit form. Different from Fourier feature encoding, the Chebyshev feature map has recently gathered attention as it generates quantum states with amplitudes proportional to Chebyshev polynomials of the first kind, forming an orthonormal Chebyshev basis on non-equidistant nodes~\cite{williams2023quantum,paine2023physicsinformed}.
%The explicit model requires a fix unitarry transformation for sampling via projective measurements. Different from the complex-valued Fourier feature encoding, the states prepared by the Chebyshev feature map are orthonormal with real-valued amplitudes on non-equidistant Chebyshev nodes.%
%

In this paper, we propose an orthonormal Hartley feature map that enables explicit quantum model building in the Hartley basis~\cite{edeclaration}. Unlike DFT, the discrete Hartley transform (DHT) maps a real input to a real output and has the convenient property of being its own inverse~\cite{Bracewell1999}, as well particularly suitable for physics-informed quantum machine learning \cite{paine2023physicsinformed}. The DHT has been shown to offer computational advantages over DFT in applications of power spectrum and convolution computations~\cite{Alexander2022,Keith2022}. The orthonormal Hartley feature map as a quantum circuit parameterized by a continuous variable $x$ prepares quantum states with amplitudes proportional to the so-called Hartley kernel function, which facilitates building quantum models in the real-valued Hartley space with an exponentially large basis set and allows for model differentiation. We apply the developed quantum protocols for learning probability distributions motivated by financially-relevant processes. We demonstrate the efficient sampling of the Hartley-based quantum model by mapping the Hartley basis to the computational basis via a quantum Hartley transform (QHT) circuit. We then employ the developed tools to solve differential equations and compare the results obtained from the different quantum models constructed by Fourier and Hartley bases. Finally, we proceed to extend and implement the proposed strategies to multivariate encoding and sampling for multidimensional quantum generative modeling.

\section*{Results and Discussion}
\subsection{I. Orthonormal Hartley feature map circuit}
We design Hartley feature maps to facilitate the learning process for quantum probability distribution, and make models trainable, while allowing easy sampling from the trained model via a unitary transformation to the computational basis. The computational basis states refer to orthonormal states $\{ |x_j\rangle \}_{j=0}^{2^N-1}$ such that overlaps equate to the Kronecker delta function, $\langle x_{j'}|x_{j}\rangle = \delta_{j',j}$. In the case where the initial input states are $|\mathrm{\o}\rangle$, $|x_j\rangle$ can be easily generated by applying a Pauli $\hat{X}$ gate to the $l^{\text{th}}$ qubit if the $l^{\text{th}}$ classical bit is 1, often referred to as the basis encoding \cite{schuld2021supervised}. Specifically, we want to generate a $N$-qubit quantum state $|h(x)\rangle$ with amplitude proportional to an equally weighted sum of $x$-dependent cosine and sine wave functions, namely the Hartley kernel $\mathrm{cas}(2\pi kx/2^N) \equiv \cos(2\pi kx/2^N)+\sin(2\pi kx/2^N)$. This state can be written as $|h(x)\rangle = 2^{-N/2} \sum_{k=0}^{2^N-1} \mathrm{cas}(2\pi kx/2^N)|k\rangle$, where $\{ |k\rangle \}$ are $2^N$ computational basis states. For $x_{j} \in [0, \cdots,2^N-1]$, the $\mathrm{cas}(\cdot)$ function satisfies the following orthogonality conditions,
%\begin{equation}
%\label{eq:h(x)}
    %|h(x)\rangle = \frac{1}{2^{N/2}} \sum_{k=0}^{2^N-1} \mathrm{cas}(2\pi kx/2^N)|k\rangle,
%\end{equation}
%%%
\begin{figure}[t]
\begin{center}
\includegraphics[width=1.0\linewidth]{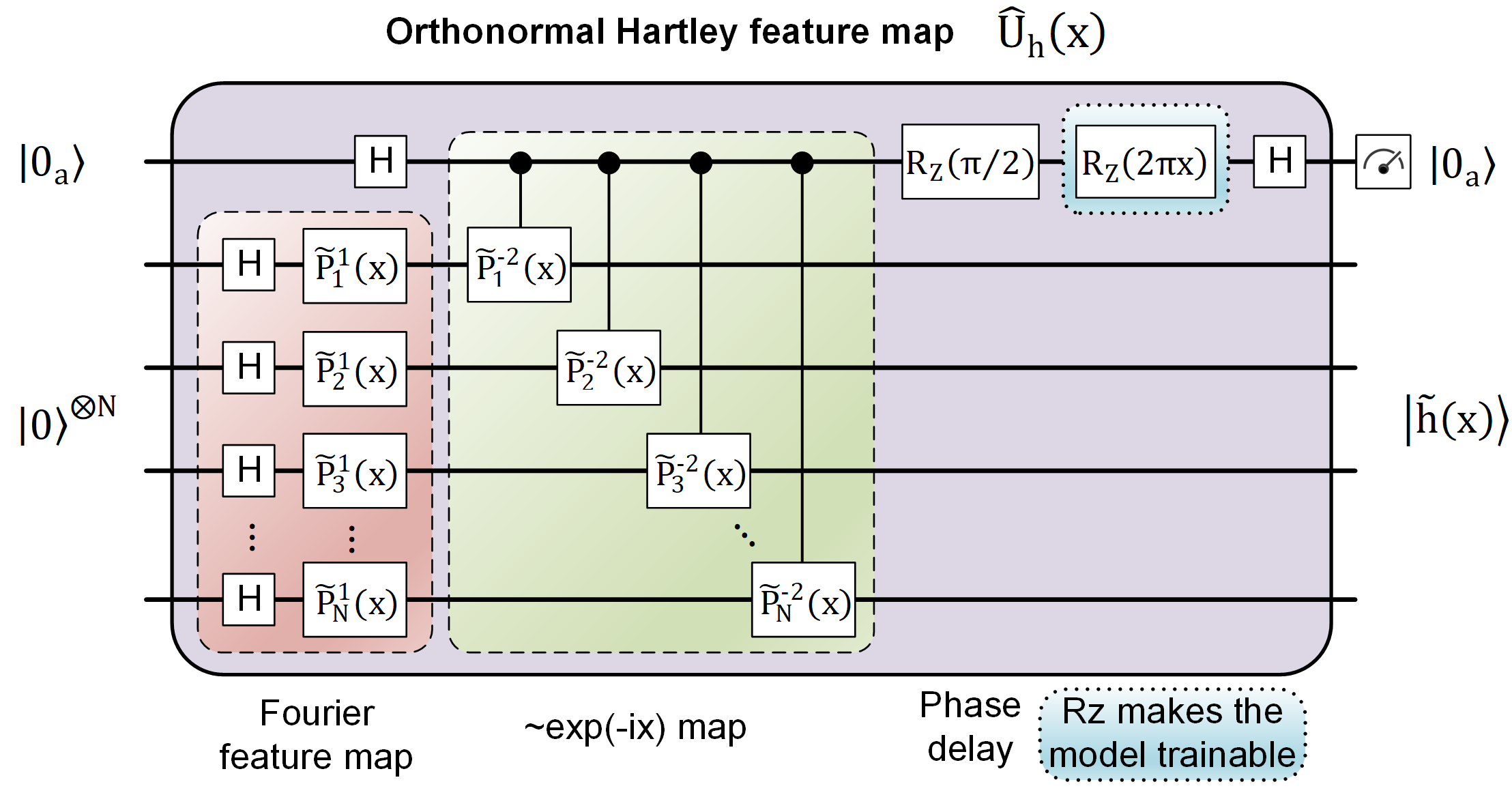}
\end{center}
\caption{\textbf{Quantum Hartley feature map.} \textbf{(a)} Quantum Hartley feature map $\hat{\mathcal{U}}_{h}(x)$ circuit that creates a Hartley state via $x$-parameterized isometry --- a $N$-qubit Fourier feature map followed by controlled phase shift gates to embed complex exponents and a constant $\mathrm{R}_\mathrm{Z}$ gate to provide a global phase delay. A $x$-dependent $\mathrm{R}_\mathrm{Z}$ gate is acting on the ancilla and sandwiched between two Hadamard gates to ensure favorable mid-point behavior while retaining the real amplitude of $|\tilde{h}(x)\rangle$ after the ancilla measurement yields $|0\rangle$ outcome. Here, scaled single-qubit phase shift gate is defined as $\tilde{\mathrm{P}}^{m}_l(x)=\mathrm{diag}\{1,\exp(i m 2\pi x /2^l)\}$, where $l \in [1,\cdots,N]$ is the qubit index and $m$ takes values of $1$ and $-2$, for any continuous variable $x \in \mathbb{R}_{2^N-1}$.}
\label{fig:Uh(x)}
\end{figure}
%%%
\begin{equation}
\label{eq:cas_properties}
    \sum_{j=0}^{2^N-1} \mathrm{cas}(2\pi kx_j/2^N) \: \mathrm{cas}(2\pi\ell x_j/2^N) = 
    \begin{cases} 2^N ~ &k = \ell, \\
    0 ~ &k \neq \ell, \\   
    \end{cases}
\end{equation}
As a consequence, the states $|h(x)\rangle$ are orthonormal on the integer grid points. Namely, the set of Hartley states $\{|h(x_{j})\rangle\}_{j=0}^{2^N-1}$ satisfies $\langle h(x_{j'})|h(x_{j})\rangle = \delta_{j',j}$ with $\delta_{j',j}$ being the Kronecker delta function. We note that the states $|h(x)\rangle$ also fulfill this orthonormal condition for half-integer points, $\{x_{(j+1/2)}\} ~\forall j \in[0, 2^N-1]$. Since the $\mathrm{cas}(\cdot)$ function can be alternatively expressed as a delayed cosine function, $\mathrm{cas}(x) = \sqrt{2}\cos(x-\pi/4)$, Hartley states can be prepared using a combination of exponents $\cos(x) = [\exp(i x) + \exp(-i x)]/2$ for some scaled variable $x$, where each amplitude is embedded via the phase feature map \cite{Kyriienko2022}.
%%%
\begin{figure*}[t]
\begin{center}
\includegraphics[width=1.0\linewidth]{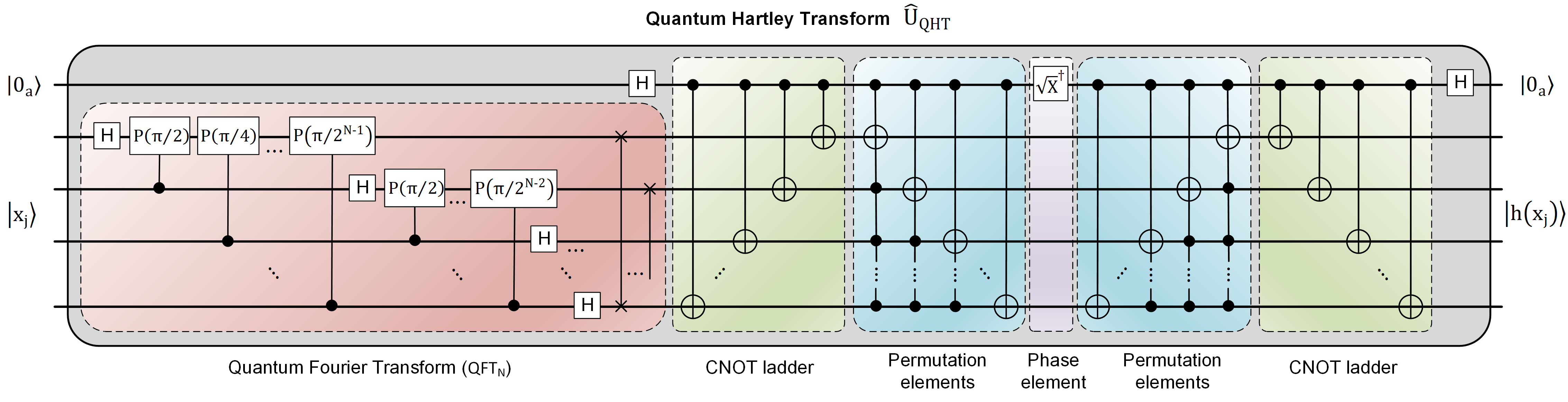}
\end{center}
\caption{\textbf{Quantum Hartley Transform.} Quantum Hartley transform $\hat{\mathcal{U}}_{\mathrm{QHT}}$ circuit which maps computational basis states $\{|0_a x_j\rangle \}_{j=0}^{2^N-1}$ into Hartley states $\{|0_a h(x_j)\rangle \}_{j=0}^{2^N-1}$, where $x_j \in \mathbb{Z}_{2^N-1}$. The QHT circuit involves a QFT circuit acting on $N$-qubit system beneath an ancilla qubit, followed by two sequential sets of CNOT ladder and permutation circuits with an inverse $\sqrt{\mathrm{X}}$ gate in the middle, sandwiched between two Hadamard gates applied to the ancilla. The standard single-qubit phase shift gate in QFT circuit is defined as $\mathrm{P}(\phi)=\mathrm{diag}\{1,\exp(i\phi)\}$. H and X are Hadamard and Pauli $\hat{X}$ gates, respectively, and $\sqrt{\mathrm{X}}^{\dagger}$ is equivalent to $\mathrm{R_Z}(\pi/2)\mathrm{P}(-\pi/2)\mathrm{R_X}(-\pi/2)$.
}
\label{fig:UQHT}
\end{figure*}
%%%
Therefore, the implementation of Hartley feature map circuit $\hat{\mathcal{U}}_{h}(x)$ that prepares normalized Hartley state for any continuous variable $x$ can be achieved by the linear combination of unitary (LCU) approach~\cite{Childs2017}, as shown in Fig.~\ref{fig:Uh(x)}. 
The circuit begins with a Hadamard gate acting on the ancilla register, being the most significant bit, with a Fourier feature map applied to $N$ qubits to distribute the scaled relative phases $\exp(i x)$ to all of the states in Fourier space, and then a series of controlled phase gates are applied to contribute relative phases based on $\exp(-i x) = \exp(i x) \exp(-i 2x)$. A constant $\mathrm{R}_\mathrm{Z}(\pi/2)$ gate is utilized to produce a $\pi/4$ phase delay in the cosine argument. A $x$-dependent $\mathrm{R}_\mathrm{Z}(2 \pi x)$ gate is appended to the ancillary qubit to minimize the contributions of the quantum state associated with non-integer points on the state overlap (see supplementary Fig.~\ref{fig:Overlap} for detailed discussion), followed by a Hadamard gate. Under the condition that the ancilla register collapses to $|0\rangle$ outcome, the quantum state prepared by $\hat{\mathcal{U}}_{h}(x)$ reads $|\tilde{h}(x)\rangle = |h(x)\rangle / \mathcal{N}(x)$ , where the unnormalized state $|h(x)\rangle$ is expressed as
\begin{equation}
\label{eq:newh(x)}
    |h(x)\rangle = \frac{1}{2^{N/2}} \sum_{k=0}^{2^N-1} \mathrm{cas} \left[ (2\pi k/2^N-\pi)x \right] |k\rangle,
\end{equation}
and $\mathcal{N}(x) = \sqrt{\langle h(x)|h(x) \rangle} = \sqrt{1-\text{sin}(2 \pi x)/2^{N}}$ is $x$-dependent when not evaluated at the half-integer and integer grid points.

% With Eq.~\eqref{eq:newh(x)}, one can construct a matrix with $2^N$ columns, where each column is represented by $|h(x_j)\rangle$. This square matrix defined as $\hat{\mathcal{U}}_{\text{H}} \coloneqq \Big\{ |h(x_0)\rangle,|h(x_1)\rangle,\cdots,|h(x_{2^N-1})\rangle \Big\} $ satisfies the matrix polynomial equation $\hat{\mathcal{U}}_{\text{H}}^4 = \hat{\text{I}}$, operating real-valued (non-involutory) $\hat{\mathcal{U}}_{\text{H}}$ four times gives back the identity matrix and in a sense being analogous to $\hat{\mathcal{U}}_{\mathrm{QFT}}^4 = \hat{\text{I}}$.%
%where the complex-valued $\hat{\mathcal{U}}_{\mathrm{QFT}}$ is the quantum Fourier transform%.

\subsection{II. Quantum Hartley transform circuit}
For sampling purposes, we need to develop a corresponding transformation circuit that enables the mapping between Hartley states $\{|h(x_j)\rangle\}_{j=0}^{2^N-1}$ %($=\{|\tilde{h}(x_j)\rangle\}_{j=0}^{2^N-1}$) 
and the computational states $\{|x_j\rangle \}_{j=0}^{2^N-1}$ (and reverse). This unitary transformation reads $|h(x_j)\rangle = (-1)^j \hat{\mathcal{U}}_{\mathrm{QHT}} |x_j\rangle$, where $\hat{\mathcal{U}}_{\mathrm{QHT}}$ represents quantum Hartley transform (QHT) and the sign flips for odd integers results from the introduction of $x$-dependent $\mathrm{R}_\mathrm{Z}$ gate in the Hartley feature map circuit. For general sampling, the effect of this phase modulation on projective measurements can be ignored. As a result, the matrix representation of QHT can be expressed as
$\hat{\mathcal{U}}_{\mathrm{QHT}} = \sum_{j=0}^{2^N-1} |h(x_j) \rangle \langle x_j|$. We note that QHT is the quantum analogue of DHT. Namely, the vector amplitude of $|h(x_j)\rangle$ corresponds to the $(j+1)^{\text{th}}$ column of the DHT matrix defined as $\mathrm{DHT}_{N} \coloneqq 2^{-N/2}\Big\{\mathrm{cas}(2\pi kj/2^N)\Big\}$ $ ~\forall$ $k,j \in [0, 2^N-1]$. It is worth mentioning that $\hat{\mathcal{U}}_{\mathrm{QHT}}$ is an involutory matrix $\hat{\mathcal{U}}_{\mathrm{QHT}}^2 = \hat{\mathcal{U}}_{\mathrm{QHT}}^{2\dagger} = \hat{\text{I}}$ in contrast to $\hat{\mathcal{U}}_{\mathrm{QFT}}^2 = \hat{\mathcal{U}}_{\mathrm{QFT}}^{2\dagger} \neq \hat{\text{I}}$. We realize that $\hat{\mathcal{U}}_{\mathrm{QHT}}$ is strongly related to the $\hat{\mathcal{U}}_{\mathrm{QFT}}$ and thus suggest the use of an extended QFT circuit to build QHT circuit~\cite{klappenecker_2001}, as shown in Fig.~\ref{fig:UQHT}. The circuit begins with a Hadamard gate applied to the ancilla, being the most significant bit, and a QFT circuit applied to $N$ qubits, which maps the input binary state $|x_j\rangle$ from the computational basis to Fourier basis. The combination of a CNOT ladder and a permutation circuit is used to reorder the amplitudes of the conditioned states and is equivalent to a controlled $\hat{\mathcal{U}}_{\mathrm{QFT}}^2$. A Hermitian adjoint of square root of Pauli $\hat{X}$ gate is introduced to adjust the relative phases and the intermediate state at this stage is in the form of $ \left( |0_a\rangle \hat{\mathcal{U}}_{\mathrm{QHT}} |x_j\rangle + |1_a\rangle \hat{\mathcal{U}}_{\mathrm{QFT}}^2 \hat{\mathcal{U}}_{\mathrm{QHT}} |x_j\rangle \right)/\sqrt{2}$. The circuit is concluded with the adjoint (conjugate transpose) version of CNOT ladder and permutation circuits, followed by a Hadamard gate to ensure that $|0_a h(x_j) \rangle$ is left alone for any input states $|0_a x_j\rangle$, and the amplitude of $|h(x_j)\rangle$ is purely real. We note that the ancilla starts and ends in $|0\rangle$ state ('clean run'). In the following sections, the symbol $\hat{\mathcal{U}}_{\mathrm{QHT}}$ will be treated as a QHT circuit rather than matrix itself, providing the unitary transformation as
\begin{equation}
\label{eq:QHT}
    \hat{\mathcal{U}}_{\mathrm{QHT}} = \sum_{j=0}^{2^N-1} |0_a h(x_j) \rangle \langle 0_a x_j|.    
\end{equation}
We also note that QHT circuit is not unique, and can be potentially optimized or recompiled for any quantum computing architecture to be used.
% The relationship between involutory $\hat{\mathcal{U}}_{\mathrm{QHT}}$ and non-involutory $\hat{\mathcal{U}}_{\text{H}}$ is $\hat{\mathcal{U}}_{\text{H}} = \hat{\mathcal{U}}_{\mathrm{QHT}} (\hat{\text{I}}^{\otimes N} \otimes \hat{Z})$.%
%
\begin{figure*}[t]
\begin{center}
\includegraphics[width=1.0\linewidth]{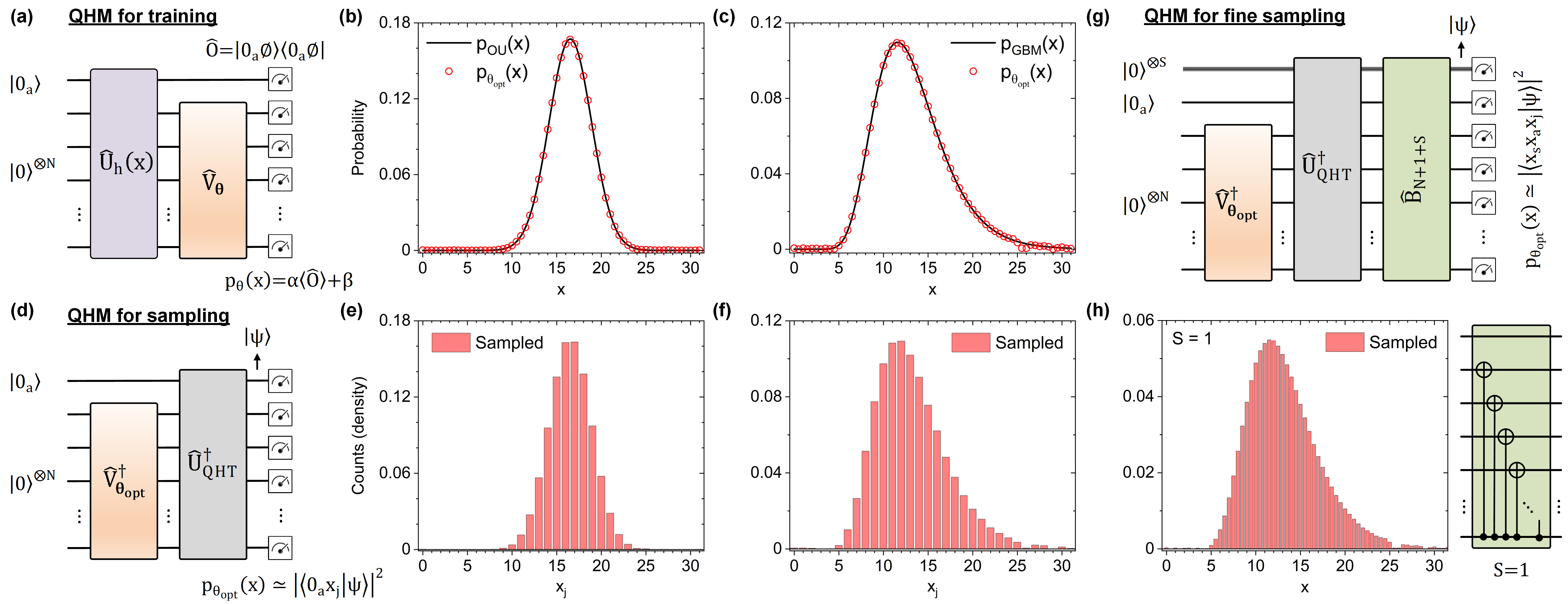}
\end{center}
\caption{\textbf{Learning and sampling probability distributions with quantum Hartley-based models (QHMs).} \textbf{(a)} Circuit used to train the model $p_{\theta}(x)$ in the latent space. Measured observable is defined as $\hat{\mathcal{O}} = |0_a \mathrm{\o}\rangle \langle 0_a \mathrm{\o}|$, where $|\mathrm{\o}\rangle \equiv |0\rangle^{\otimes N}$, with $\alpha$ and $\beta$ being trainable scaling and bias parameters.  \textbf{(b)} Trained $p_{\theta_\text{opt}}(x)$ and target $p_{\text{OU}}(x)$ distributions with parameters $\mu=5$, $\sigma=3$, $\nu=0.5$, $x_i=24$ and $t=1$. \textbf{(c)} Trained $p_{\theta_\text{opt}}(x)$ and target $p_{\text{GBM}}(x)$ distributions with parameters $\mu=0.1$, $\sigma=0.3$, $x_i=12$ and $t=1$ for $x > 0$.  \textbf{(d)} Circuit used to sample the trained model in the computational basis $|0_ax_j\rangle$. The quantum state just prior to measurement is denoted as $|\psi\rangle$. \textbf{(e, f)} Sampled probability distributions from the corresponding trained models, generated with $10^5$ shots. The normalized histogram is plotted with respect to $x_j \in \mathbb{Z}_{2^N-1}$. \textbf{(g)} Circuit with $S$-qubit extended registers on the top line used to perform fine sampling in the computational basis $|x_sx_ax_j\rangle$. \textbf{(h)} Sampled probability distribution from the trained (GBM) model using the extended register of $S = 1$, generated with $10^6$ shots. The normalized histogram is plotted as a function of $x \in \mathbb{R}_{2^N-1}$. The bitstring network $\hat{\mathcal{B}}$ for $S=1$ is shown on the right.} 
\label{fig:SDEresults}
\end{figure*}
%%%

\subsection{III. Learning and sampling}
Next, we demonstrate examples of applying the quantum Hartley transform to generative modeling of two relevant distributions of stochastic models corresponding Ornstein-Uhlenbeck (OU) and geometric Brownian motion (GBM) processes, which among others are widely used in financial analysis. For instance, the former is typically employed to model interest rates and currency exchange rates~\cite{VASICEK1977177}, while the latter is usually utilized to model the log return of stock prices in the Black–Scholes model~\cite{black_scholes}. The stochastic process $X_t$ at time $t$ satisfies a stochastic differential equation (SDE) $dX_t = \nu(\mu-X_t)dt+\sigma dW_t$ for OU and $dX_t=\mu X_tdt + \sigma X_tdW_t$ for GBM process, where the constant parameters ($\mu$, $\sigma$ and $\nu$) represent mean (drift), volatility and reversion speed, respectively, and $W_t$ denotes the Wiener process. Using the Fokker-Planck equation to treat $X_t$ as a deterministic variable $x$, we obtain the time evolution of the underlying probability density function $p(x,t)$ for a given initial distribution of $p(x,0) =\delta(x-x_i)$. The results show that $X_t$ follows normal and log-normal distributions, respectively, as
\begin{equation}
\label{eq:OU}
    p_{\text{OU}}(x,t) = \sqrt{\frac{\nu}{\pi (1-\textit{e}^{-2\nu t})\sigma^2}}\text{exp} \left\{ \frac{-\nu \left[ x-\mu-(x_i-\mu)\textit{e}^{-\nu t} \right]^2}{(1-\textit{e}^{-2\nu t})\sigma^2} \right\},
\end{equation}
\begin{equation}
\label{eq:GBM}
    p_{\text{GBM}}(x,t) = \frac{1}{\sqrt{2\pi \sigma^2 t}\ x}\text{exp} \left\{ \frac{- \left[ \text{ln}(x/x_i)-(\mu-\sigma^2/2)t \right]^2}{2\sigma^2t} \right\},
\end{equation}
Setting $p(x,t\rightarrow 1) = p_{\text{target}}(x)$ as a target distribution (ground truth), we make use of the Hartley feature map followed by an ansatz parameterized by a vector of variational parameters $\bm{\theta}$ that can be adjusted in a hybrid classical-quantum optimization scheme to learn these two distributions via the differentiable quantum generative model framework (DQGM)~\cite{Kyriienko2022}. To maintain the real amplitudes of the state vector, the variational ansatz $\hat{\mathcal{V}}_{\theta}$ comprises adjustable $\mathrm{R}_\mathrm{Y}$ rotation and fixed CNOT/CZ gates. An exemplary $\hat{\mathcal{V}}_{\theta}$ used in this study is illustrated in supplementary Fig.~\ref{fig:HERA}. Specifically, the $N$-qubit hardware efficient real-amplitude ansatz (HERA) consists of an initial layer of parameterized $\mathrm{R}_\mathrm{Y}$ gates applied on each qubit, followed by depth-$d$ blocks presented in the alternating layered architecture. Each block is composed of a cascade of entangling (CNOT) gates and a layer of parameterized $\mathrm{R}_\mathrm{Y}$ gates applied on each qubit. The total $N(d+1)$ trainable parameters are randomly initialized. This ansatz prepares highly correlated quantum states with real-only amplitude while maximizing the expressivity of the model~\cite{Schuld2020,Schuld2021}.
%where $|\mathrm{\o}\rangle = |0\rangle^{\otimes N}$

We begin with variationally training both normal and log-normal distributions in the latent space through feeding an initial product state $|0_a \mathrm{\o}\rangle$ to the Hartley feature map $\hat{\mathcal{U}}_h(x)$ connected with a variational ansatz circuit $\hat{\text{I}}\otimes \hat{\mathcal{V}}_{\theta}$, and then read out the quantum model as an expectation value of an observable $\hat{\mathcal{O}} = |0_a \mathrm{\o}\rangle \langle 0_a \mathrm{\o}|$. Note that the latter can be substituted by a local proxy during the training stage \cite{Cerezo2021NatComm}. In order to improve the trainability and expressivity of QML models, they are usually formulated as $ p_\theta(x) = \alpha \langle\hat{\mathcal{O}}\rangle + \beta $ with variationally trainable scaling and (optional) bias parameters ($\alpha$ and $\beta$), as depicted in Fig.~\ref{fig:SDEresults}(a). The quantum model is trained to search for optimized $\bm{\theta _\text{opt}}$ to fit the target probability distribution by minimizing a mean squared error (MSE) loss function
\begin{equation}
\label{eq:MSE}
   \mathcal{L}(\theta) = \frac{1}{M} \sum_{m=1}^{M} \left[p_{\theta}(x_m) - p_{\text{target}}(x_m) \right]^2,
\end{equation}
where $M$ is a grid of training points consisting of the integers $\{x_j\}$ and additional half-integers $\{x_{(2j+1)/2}\} ~\forall j\in[0, 2^N-1]$. The loss minimization is performed via Adam optimizer for gradient-based training of variational parameters $\bm{\theta}$.
%between parameterized and ground-truth distributions% 
In Fig.~\ref{fig:SDEresults}(b,c) we show the trained normal and log-normal distributions using $N$ = 5 qubits with $\hat{\mathcal{V}}_{\theta}$ of depth $d$ = 4 and 5, respectively. Trained models (red circles) tightly follow the target functions (black solid curves). Because qubits are entangled and rotated differently from the initial state to the target state in each learnable block, the number of depth blocks can significantly impact the learning accuracy of both financially-motivated models.
\begin{figure}[t]
\begin{center}
\includegraphics[width=1.0\linewidth]{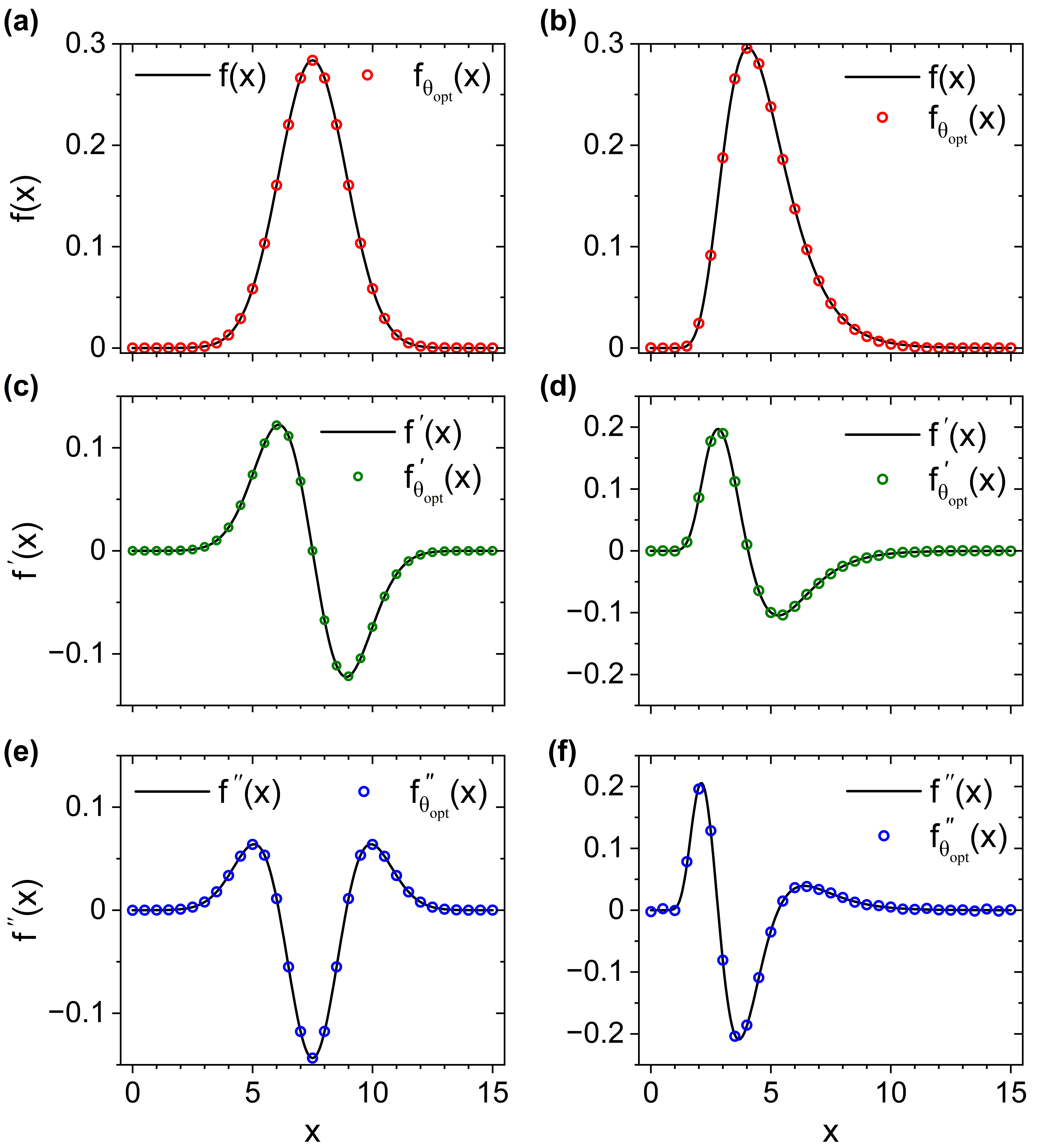}
\end{center}
    \caption{\textbf{Results of solving differential equations.} \textbf{(a)} Plots of trained model $f_{\theta_\text{opt}}(x)$ and analytic solution $f(x)$ of Eq.~\eqref{eq:DE1} with boundary conditions of $f(\mu) = 1/\sqrt{2 \pi \sigma^2}$ and $\eval{df(x)/dx}_{x=\mu} = 0$, where parameters are $\mu=7.5$ and $\sigma=1.406$.
    \textbf{(b)} Plots of trained model $f_{\theta_\text{opt}}(x)$ and analytic solutions $f(x)$ of Eq.~\eqref{eq:DE2} with boundary conditions of $f(e^{\mu-\sigma^2}) = e^{(\sigma^2/2-\mu)}/\sqrt{2 \pi \sigma^2}$ and $\eval{df(x)/dx}_{x=e^{\mu-\sigma^2}} = 0$, where parameters are $\mu=1.5$ and $\sigma=0.316$.
    The corresponding \textbf{(c, d)} first and \textbf{(e, f)} second derivatives of the analytic solutions and trained models.}
\label{fig:DEresults}
\end{figure}

Since $p_{\theta_\text{opt}}(x) = |\langle 0_a \mathrm{\o}|\bigl(\hat{\text{I}}\otimes \hat{\mathcal{V}}_{\theta_\text{opt}} \bigl) |0_a \tilde{h}(x)\rangle|^2  \simeq  |\langle 0_a x_j|\hat{\mathcal{U}}_{\mathrm{QHT}}^\dagger \bigl(\hat{\text{I}}\otimes \hat{\mathcal{V}}_{\theta_\text{opt}}^{\dagger} \bigl) |0_a \mathrm{\o}\rangle|^2$, we can sample the probability distribution of the trained model by applying the adjoint versions of the trained ansatz and QHT circuits to the zero product input state, $|\psi\rangle = \hat{\mathcal{U}}_{\mathrm{QHT}}^\dagger \bigl(\hat{\text{I}}\otimes \hat{\mathcal{V}}_{\theta_\text{opt}}^{\dagger} \bigl) |0_a \mathrm{\o}\rangle$, and then perform projective measurements in the computational basis to collect a batch of binary samples, as shown in Fig.~\ref{fig:SDEresults}(d). The histograms in Fig.~\ref{fig:SDEresults}(e,f) show the resulting sampled probability distributions normalized with the total number of samples, and both of them are in good agreement with the corresponding target distributions [solid curves in Fig.~\ref{fig:SDEresults}(b,c)]. One can readily increase the sampling rate at the expense of decrease of maximum probability amplitude through $S$-qubit extended registers on the top line, $x_s \in [0,\cdots,2^S-1]$. The relevant circuits associated with the extended registers are ($N$+1+$S$)-qubit inverse QHT and bitstring network $\hat{\mathcal{B}}$ circuits before projective measurements, as depicted in Fig.~\ref{fig:SDEresults}(g). The requirement of the inclusion of an extra bitstring network for $S \geq 1$ originates from the use of $x$-dependent R$_Z$ gate acting on the ancilla in the Hartley feature map during the training procedure. The nature of $x$-dependent argument makes the training stable at the cost of having (bit-dependent) periodic signal on the ancilla. As a demonstration of the double-frequency sampling, the histogram of the sampled probability distribution from the trained (GBM) model with the extended register of $S$ = 1 is plotted as a function of $x \in \mathbb{R}_{2^N-1}$ and presented in Fig.~\ref{fig:SDEresults}(h), where the length-($N$+1+$S$) readout binary strings are linearly mapped to the domain $x$ in the range of $[0, 2^{N+1}-1]$. We note that the bitstring network is just a CNOT ladder for $S$ = 1. Further fine sampling can be achieved by increasing extended register sizes together with different types of bitstring network. The other way to perform dense sampling is borrowed from the sampling procedure of Fourier model. One can transform the computational states to Fourier states by a ($N$+1)-qubit QFT circuit, followed by an extended inverse QFT circuit, as shown in supplementary Fig.~\ref{fig:SFig3}, where the histograms for double- and quadruple-frequency sampling corresponding to different sizes of the extended registers are demonstrated. More importantly, the resulting histograms from both sampling configurations qualitatively match the target distribution, providing a potential pathway to generate unseen datasets associated with those untrained grids or draw samples from a parameterized probability distribution for applications of quantum generative models in machine learning tasks and quantum physics. In the following sections, we will employ the developed protocols on solving differential equations, making a comparison between Fourier and Hartley models, and exploring the possibility toward multidimensional quantum generative modeling.
%followed by a $\hat{\text{I}}^{\otimes S} \otimes$ $(N+1)-$qubit QFT circuit and a $(N+1+S)-$qubit inverse QFT circuit before projective measurements, as depicted in Fig.~\ref{fig:SDEresults}(g).
%$\hat{\mathcal{U}}_{\mathrm{QHT}}^\dagger \bigl(\hat{\text{I}}\otimes \hat{\mathcal{V}}_{\theta ^\ast}^{\dagger} \bigl) |\mathrm{\o}\rangle$
%\textit{e}^{-0.5((x-\mu)/\sigma)^2}
%

\subsection{IV. Solving differential equations}
We first consider two exemplary second-order differential equations (DEs) with variable coefficients to be tackled. The first DE is of the form 
\begin{equation}
\label{eq:DE1}
     \frac{d^2f(x)}{dx^2} + \frac{(x-\mu)}{\sigma^2} \frac{df(x)}{dx} + \frac{ f(x)}{\sigma^2} = 0, 
     %\\ f(\mu) = \frac{1}{\sqrt{2 \pi}\,\sigma}, \quad
     %\eval{\frac{df(x)}{dx}}_{x=\mu} = 0
\end{equation}
for some real-valued parameters $\mu$, $\sigma$ and the boundary conditions $f(\mu) = 1/\sqrt{2 \pi \sigma^2}$ and $\eval{df(x)/dx}_{x=\mu} = 0$. This differential equation has a known analytical solution $ f(x)= \text{exp} \left\{-0.5 \left[ (x-\mu)/\sigma \right]^2 \right\}/\sqrt{2 \pi \sigma^2} $. The second DE reads
\begin{equation}
\label{eq:DE2} 
    \frac{d^2f(x)}{dx^2} + \frac{[2\sigma^2-\mu+\text{ln}(x)]}{\sigma^2x} \frac{df(x)}{dx} +  \frac{f(x)}{\sigma^2x^2} = 0, 
    %\\f(e^{\mu-\sigma^2}) =  \frac{e^{\frac{\sigma^2}{2}-\mu}}{\sqrt{2\pi}\,\sigma}, \quad  
    %\eval{\frac{df(x)}{dx}}_{x=e^{\mu-\sigma^2}} = 0
\end{equation}
with the boundary conditions $f(e^{\mu-\sigma^2}) = e^{(\sigma^2/2-\mu)}/\sqrt{2 \pi \sigma^2}$ and $\eval{df(x)/dx}_{x=e^{\mu-\sigma^2}} = 0$. The analytical solution of this DE is $f(x)=\text{exp} \left\{-0.5 \left[ (\text{ln}x-\mu)/\sigma \right]^2 \right\}/(\sqrt{2 \pi \sigma^2} \, x)$ for $x > 0$. 
Solving DEs requires evaluations of the first and second derivatives of the quantum model $ f_{\bm{\theta}}(x) = \alpha \langle 0_a \mathrm{\o}| \hat{\mathcal{U}}_{h}^{\dagger}(x) \bigl(\hat{\text{I}}\otimes \hat{\mathcal{V}}_{\theta}^{\dagger} \bigl)   \hat{\mathcal{O}} \bigl(\hat{\text{I}}\otimes \hat{\mathcal{V}}_{\theta} \bigl)  \hat{\mathcal{U}}_{h}(x)  |0_a \mathrm{\o}\rangle + \beta$ with respect to $x$, which relies on a differentiable Hartley feature map $\hat{\mathcal{U}}_{h}(x)$ circuit. Gradient calculations are implemented with automatic differentiation techniques (backpropagation) or with the application of the parameter shift rule~\cite{schuld2019evaluating,mitarai2018quantum,kyriienko2021generalized}. For the latter case, controlled gates $\tilde{\mathrm{P}}^{-2}_l(x)$ need to be decomposed using CNOT conjugations, and differentiation of $\hat{\mathcal{U}}_{h}(x)$ requires $4N+2$ shifts in total. We then variationally minimize an overall MSE loss function. Specifically, the total loss function is the equal-weighted sum of the contributions that match the differential equation and satisfy the boundary conditions. The loss is minimized with the gradient-based Adam optimizer with a small learning rate of 0.01 and generally the loss error converges to the level of $10^{-6}$ in a few thousand epochs. We use the Python programming language (PennyLane from Xanadu \cite{bergholm2022pennylane} and Qadence from Pasqal \cite{qadence2024pasqal}) together with machine learning packages (NumPy, JAX and PyTorch) for the full statevector simulation. In particular, we utilize Hartley encoding over $N=4$ qubits and the HERA of depth 3 to solve Eqs.~\eqref{eq:DE1} and~\eqref{eq:DE2}. The results are presented in Fig.~\ref{fig:DEresults}(a)-(f). The red, green, and blue hollow-circle curves represent functions and their first and second derivatives, respectively, $f_{\theta_\text{opt}}(x)$, $df_{\theta_\text{opt}}(x)/dx$ and $d^2f_{\theta_\text{opt}}(x)/dx^2$ evaluated at optimal angles $\theta_\text{opt}$ retrieved after the optimization procedure. Overall, the trained models are consistent with the analytic solutions (solid black curves).
%%%
\begin{figure}[t]
\begin{center}
\includegraphics[width=1.0\linewidth]{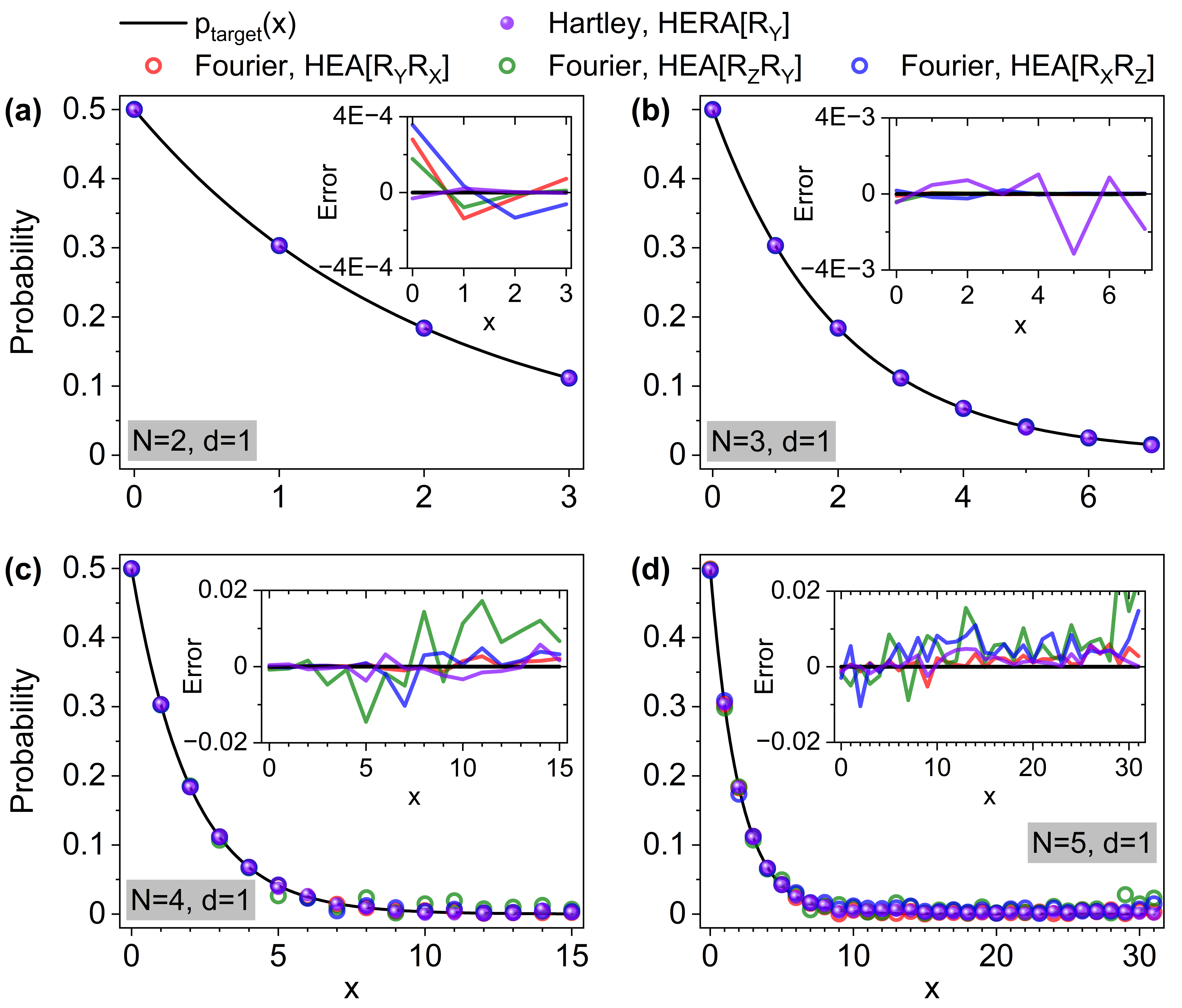}
\end{center}
    \caption{\textbf{Comparison between the quantum Fourier and Hartley models.} Training a probability exponential distribution with scale inversely proportional to parameter $\lambda$ = 0.5, $ p_{\text{target}}(x) = \lambda e^{-\lambda x} $ for $x \geq 0$, using Fourier and Hartley feature map circuits over $N$ = 2 \textbf{(a)}, 3 \textbf{(b)}, 4 \textbf{(c)}, and 5 qubits \textbf{(d)} with the shallow ansatz circuit depth of $d$ = 1. For Fourier models, the parameterized gate element in the HEA is composed of two adjustable single-qubit rotation gates per qubit $(\mathrm{R}_\mathrm{Y}\mathrm{R}_\mathrm{X}$, $\mathrm{R}_\mathrm{Z}\mathrm{R}_\mathrm{Y}$, $\mathrm{R}_\mathrm{X}\mathrm{R}_\mathrm{Z}$, expressed in matrix multiplication order). For Hartley models, only the $\mathrm{R}_\mathrm{Y}$ gate is required in the HERA. The insets show the corresponding relative errors.
    }     
\label{fig:Comparisions}
\end{figure}
%%%

\subsection{V. Complex Fourier vs real Hartley models}
To evaluate the potential of using the Hartley models, we consider an exponential distribution with probability density function $ p(x) = \lambda e^{-\lambda x} $ as a ground truth. This $p(x)$ describes the distances between successive events in homogeneous Poisson processes. As being the only continuous memoryless probability distribution, it is widely used in the calculation of various systems in queuing theory and reliability theory~\cite{Feller}. We employ $N$-qubit Fourier encodings followed by different configurations of hardware efficient ansatz (HEA) to learn this exponential distribution on a training grid of $2^N$ integer points and compare the results with Hartley models. Note that the ancillary register is not required for Fourier models [see Fig.~\ref{fig:Overlap}(c)]. Due to the complex nature of the Fourier models, the parameterized gate element inside each depth block of the HEA is composed of two single-qubit rotation gates per qubit (marked as $\mathrm{R}_\mathrm{Y}\mathrm{R}_\mathrm{X}$,  $\mathrm{R}_\mathrm{Z}\mathrm{R}_\mathrm{Y}$, $\mathrm{R}_\mathrm{X}\mathrm{R}_\mathrm{Z}$ in Fig.~\ref{fig:Comparisions}), each gate parameterized by a given angle, and the CNOT-based entangling layers are the same as the HERA (see supplementary Fig.~\ref{fig:SFig4} for other cases of parameterized gates). Therefore, there are $2N(d+1)$ trainable parameters for HEA of depth $d$. These HEAs allow easy access to the solution space by preparing quantum states with complex amplitude and avoid the training issues such as barren plateaus and local minima for complex Fourier models. As shown in Fig.~\ref{fig:Comparisions}(a-d), both Fourier and Hartley models follow $ p_{\text{target}}(x)$ closely with small relative errors for varying number of qubits $N$ under the same ansatz depth $d$ = 1. However, the number of variational parameters in complex Fourier models is double compared to those in real Hartley models. Our results indicate for a $N$-qubit quantum system ($N \geq 2$) that one can simply employ the Hartley model with the minimal number of parameterized gates needed to efficiently reach the target solution space compared to the Fourier model. This becomes important when considering practical implementations with limited quantum resources and relevant applications.
%%%
\begin{figure}[t]
\begin{center}
\includegraphics[width=1.0\linewidth]{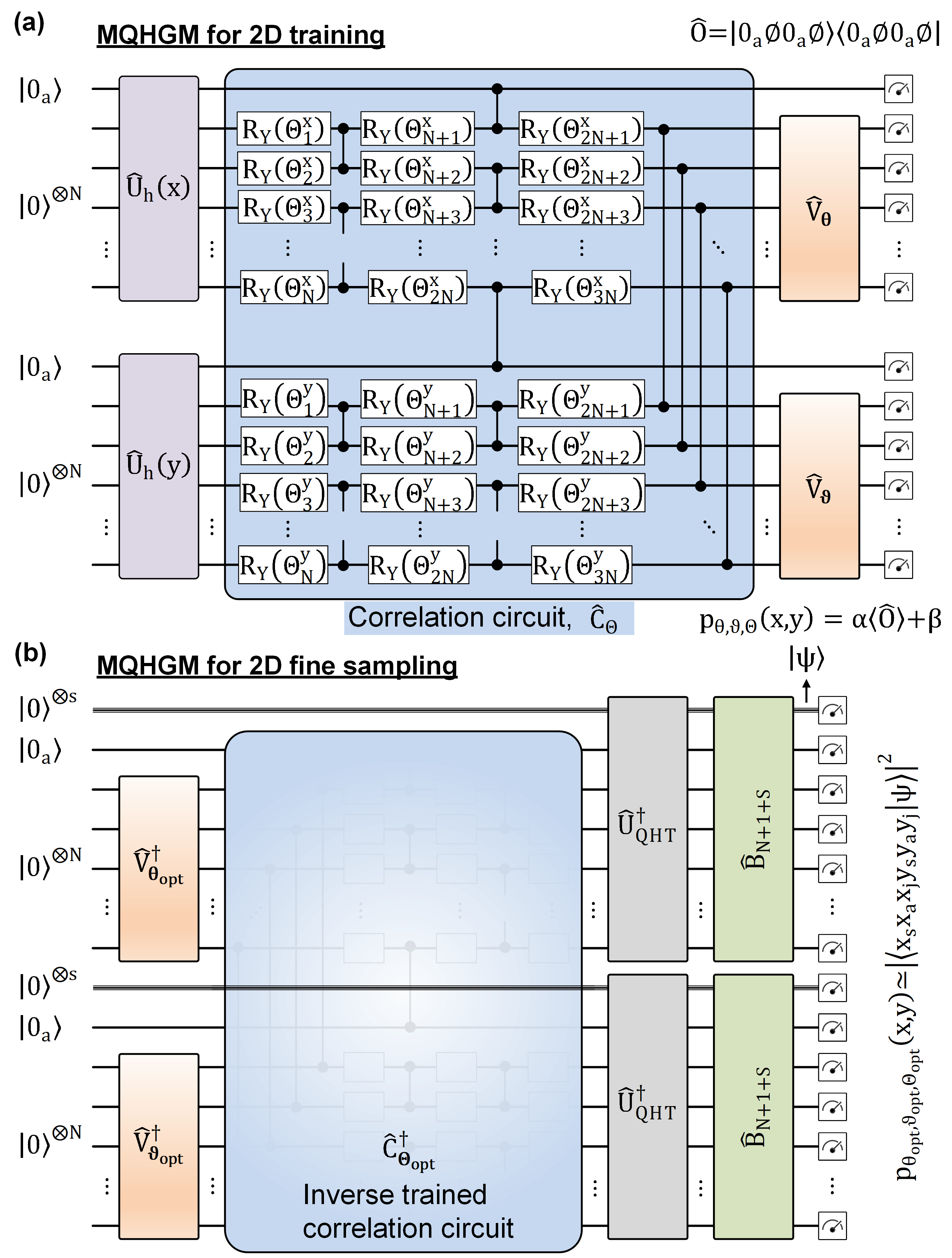}
\end{center}
    \caption{
    \textbf{Multivariate quantum Hartley-based generative models (MQHGMs).}
    \textbf{(a)} Circuit used to train the multivariate distribution in the latent space, where a parameterized correlation circuit $\hat{\mathcal{C}}_{\Theta}$ is sandwiched between feature maps ($\hat{\mathcal{U}}_{h}(x)$ and $\hat{\mathcal{U}}_{h}(y)$) and variational ansatz ($\hat{\mathcal{V}}_{\theta}$ and $\hat{\mathcal{V}}_{\vartheta}$). Measured observable is defined as $\hat{\mathcal{O}} = |0_a \mathrm{\o} 0_a \mathrm{\o} \rangle \langle 0_a \mathrm{\o} 0_a \mathrm{\o}|$ is utilized, where $|\mathrm{\o}\rangle \equiv |0\rangle^{\otimes N}$. Here, $\alpha$ and $\beta$ are trainable scaling and bias parameters. \textbf{(b)} Circuit used to sample the multivariate distribution from the trained model, where  $\bm{\theta _\text{opt}}$, $\bm{\vartheta _\text{opt}}$ and $\bm{\Theta _\text{opt}}$ are retrieved after the optimization procedure in \textbf{(a)}. Two identical sets of inverse QHT ($\hat{\mathcal{U}}_{\mathrm{QHT}}^\dagger$) and bitstring network ($\hat{\mathcal{B}}$) circuits associated with extended registers of $S$ qubits are then applied in parallel for fine sampling in the computational basis $|x_sx_ax_jy_sy_ay_j\rangle$. The quantum state just prior to measurement is denoted as $|\psi\rangle$. 
    }     
\label{fig:2DQCs}
\end{figure}
%%%

\begin{figure*}[t]
\begin{center}
\includegraphics[width=1.0\linewidth]{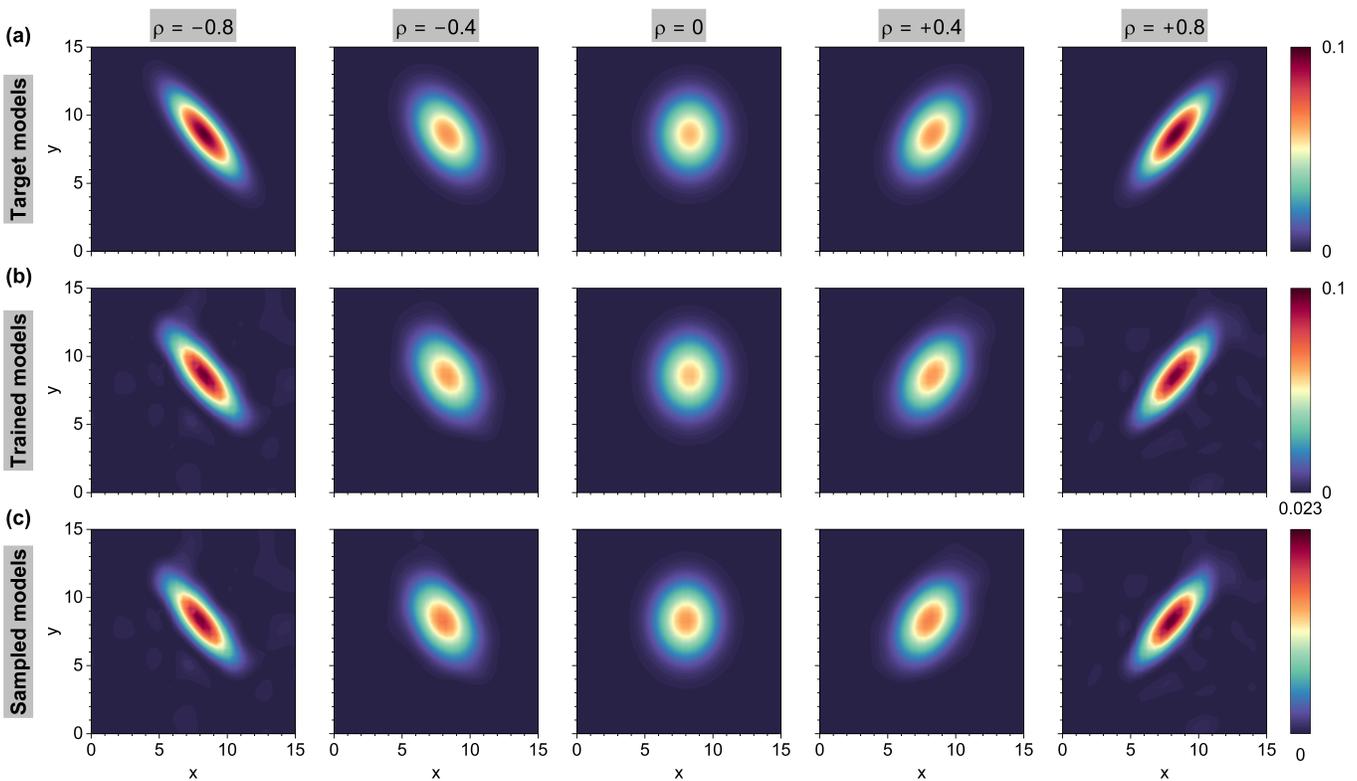}
\end{center}
\caption{
\textbf{Learning and sampling bivariate probability distributions with multivariate quantum Hartley-based generative models.} \textbf{(a)} Density plots of target $p_{\text{BN}}(x,y)$ distributions with parameters $\mu_x=8.3$, $\mu_y=8.6$ and $\sigma_x=1.5$, $\sigma_y=1.8$ for different correlation coefficients ranging from $\rho$ = $-0.8$ to $0.8$ in a step of $0.4$ (left to right). \textbf{(b)} Density plots of corresponding trained $p_{\theta_\text{opt},\vartheta_\text{opt},\Theta_\text{opt}}(x,y)$ models. \textbf{(c)} Normalized density plots of sampled distributions from the corresponding trained models in \textbf{(b)} using the parallel extended registers of \textit{S} = 1, where $10^7$ shots are measured at the readout. All plots share the same color bar on the right in each row with the same correlation coefficient in each column.
}
\label{fig:2Dresults}
\end{figure*}

\subsection{VI. Multivariate quantum generative models}
We proceed to extend the proposed strategies from univariate to multivariate distributions. As an example, we consider a bivariate/binormal distribution with probability density function in the form
\begin{equation}
\label{eq:BN}
    p_{\text{BN}}(x,y) = \frac{1}{2\pi\sqrt{1-\rho^2}\ \sigma_x \sigma_y}\text{exp} \left[ \frac{- \left(z_x^2 + z_y^2 - 2\rho z_x z_y \right)}{2 (1-\rho^2)} \right],
\end{equation}
where $z_x=(x-\mu_x)/\sigma_x$ and $z_y=(y-\mu_y)/\sigma_y$ with parameters $\mu_x$, $\mu_y$ and $\sigma_x$, $\sigma_y$ representing mean and standard deviation values for each (either stochastic or deterministic) variables $x$ and $y$. Here, $\rho$ is a correlation coefficient $-1 < \rho < 1$.  We start with Hartley encoding of two independent variables $x$ and $y$ in parallel registers, followed by a parameterized correlation circuit $\hat{\mathcal{C}}_{\Theta}$ and two separate variational ansatz $\hat{\mathcal{V}}_{\theta}$ and $\hat{\mathcal{V}}_{\vartheta}$, as illustrated in Fig.~\ref{fig:2DQCs}(a). The quantum model $p_{\theta,\vartheta,\Theta}(x,y)$ is trained to represent the target $p_{\text{BN}}(x,y)$ through searching for optimal angles, $\bm{\theta _\text{opt}}$, $\bm{\vartheta _\text{opt}}$ and $\bm{\Theta _\text{opt}}$, the same procedure in the univariate case. The correlation circuit plays a crucial role in making two otherwise independent latent variables correlated for efficient training, while keeping the sampling procedure the same as previously discussed in the univariate case. As usual, $\hat{\mathcal{C}}_{\Theta}$ comprises alternating layers of adjustable $\mathrm{R}_\mathrm{Y}$ rotation and fixed CZ gates to ensure the real amplitudes of the state vector. Specifically, the $(2N+2)$-qubit correlation circuit consists of three layers of parameterized $\mathrm{R}_\mathrm{Y}$ gates applied on each qubit (except for two ancilla), two layers of CZ gates applied to odd and even subsequent pairs of qubits, respectively, and a final layer of CZ gates applied to the same qubit index between separate registers. Therefore, the overall number if trainable parameters includes $2N(d+1)$ contributed by two variational ansatz, and $6N$ by the correlation circuit, all of which are randomly initialized. For the circuit used to densely sample multivariate distributions illustrated in Fig.~\ref{fig:2DQCs}(b), we run the trained circuit with inverted parameters in reverse order and replace the Hartley feature maps with extended inverse QHT circuits, a procedure accounting for the basis transformation from the latent to bitbasis space. Two identical bitstring networks are then applied in parallel, followed by projective measurements in the computational basis to collect a batch of binary samples, as described in the previous section. Similarly, its variant model via QFTs is shown in supplementary Fig.~\ref{fig:SFig5}.

Fig.~\ref{fig:2Dresults}(a) shows density plots of analytical bivariate (binormal) distribution based on Eqs.~\eqref{eq:BN} with five different correlation coefficients, set as 2D target distributions (ground truths). The corresponding density plots of the trained quantum models with $N$ = 4 qubits and the variational ansatz of depth $d$ = 2 are displayed in each column of Fig.~\ref{fig:2Dresults}(b), respectively. As expected, they all quantitatively follow target distributions presented in Fig.~\ref{fig:2Dresults}(a). Typically the loss error reaches to the level $10^{-6}$ after a few hundred iterations for each scenario. Successful bivariate learning over a wide range of correlation coefficients can be attributed to the well-designed architecture of the latent-space training of the quantum model with a problem-specific correlation circuit included. The proposed correlation circuit structure is designed to simultaneously support the successful training of uncorrelated, partially and highly (positive/negative) correlated distributions with $|\rho|$ smaller than 0.9. For the cases of $|\rho|$ higher than 0.9, increasing the depth of the variational ansatz circuit or a modification of the correlation circuit is essential. After projective measurements in the computational basis, we have to perform classical post-processing tasks on a batch of measured binary datasets. This involves with periodically dropping out those bits with zero probability, concatenating rest of the bits in a sequential way and reshaping the resulting bitstring in a 2D array for data plotting. For the case of the extended register of $S$ = 1, the corresponding normalized density plots of sampled distributions are shown in Fig.~\ref{fig:2Dresults}(c), which are in excellent agreement with the target distributions. In the cases of $|\rho|$ = 0.8, some defects (light-purple spots) emerge from the backgrounds of both trained and sampled models, but they do not affect the identification of appearance of the binormal distributions. Finally, the proposed framework enables us to systemically build multidimensional quantum models and easily scale to larger system sizes without the need of a huge amount of change. The size of each modular unit such as Hartley feature map, variational ansatz, quantum Hartley transform and bitstring network, is essentially determined by ``local'' Hilbert space independent of neighboring subsystem and each individual module has the same circuit architecture as that in the univariate case. That is to say, we only need to focus on the modification or redesign of layers of CZ gates responsible for ``global'' correlations among three separate registers when tackling $p(x,y,z)$ associated with three correlated independent variables. As a consequence, we envision a future research of quantum generative modeling going beyond simple univariate towards complex multivariate diffusion models using the developed techniques presented in this work.

\section*{Conclusion}
In this study, we developed protocols for building quantum machine learning models based on Hartley kernels. In these models embedding is represented by a state with amplitudes being $x$-dependent functions scaling as $\mathrm{cas}(2\pi kx/2^N)$. Being real-valued, Hartley models are suitable for solving tasks where this is required by symmetry, and provide an advantage over Fourier basis states with inherently complex coefficients. We proposed a data-dependent embedding circuit to generate the exponentially expressive orthonormal Hartley basis in the latent space, enabling the differentiation of quantum models. To support the model building, we designed a real-amplitude ansatz for efficient training of quantum models. We constructed the quantum Hartley transform circuit for mappings between Hartley and computational bases. With these tools, we performed generative modeling for probability distributions being solutions of stochastic differential equations that arise in financial modelling. We demonstrated efficient sampling of these distributions in computational basis, revealing the consistent profiles between the learnt and sampled probability distributions. We then solved the differential equations and demonstrated favorable generalization properties of Hartley-based quantum models. Finally, we showcased the capability of multivariate quantum generative modeling, where we developed a problem-specific correlation circuit and used parallel extended registers. This opens a way to multivariate sampling enabled by differentiable physics-informed models.

%\vspace{1mm}

\section*{Acknowledgements}
We thank Annie E. Paine, Chelsea A. Williams, and Antonio Andrea Gentile for useful discussions on the project.

%\printbibliography
%\bibliography{bibliography}
%

\clearpage
\newpage 
%\appendix

\newcommand{\beginsupplement}{
    \setcounter{table}{0}
    \renewcommand{\thetable}{S\arabic{table}}
    \setcounter{figure}{0}
    \renewcommand{\thefigure}{S\arabic{figure}}
}  

\beginsupplement
\section*{Supplementary information}

\begin{figure}[ht]
\begin{center}
\includegraphics[width=1.0\linewidth]{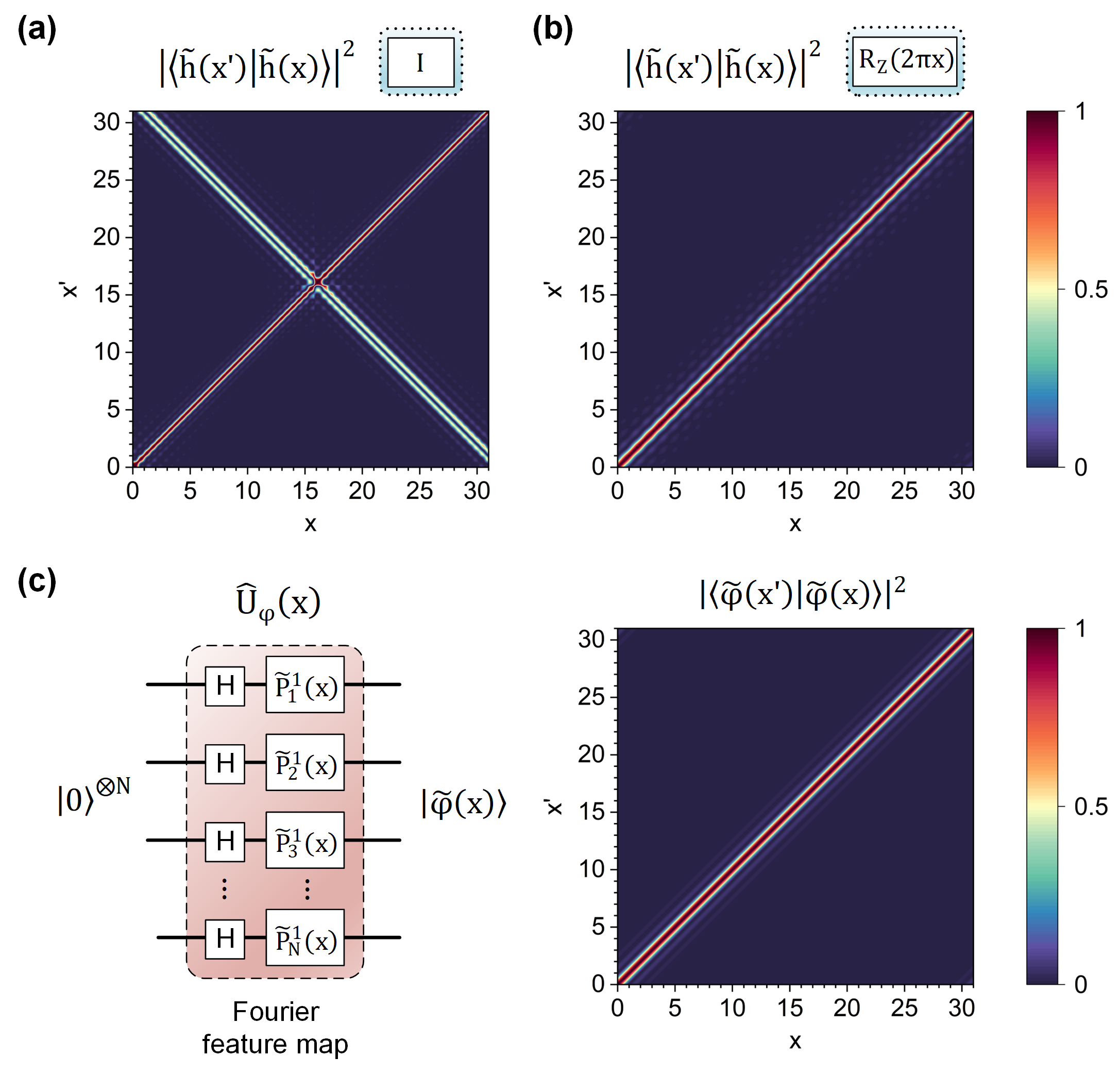}
\end{center}
    \caption{\textbf{ Two-dimensional squared overlap between mutual normalized states.}
     \textbf{(a,b)} Representative two-dimensional (2D) squared overlap between normalized Harley states $|\langle \tilde{h}(x')|\tilde{h}(x) \rangle|^2$ for $\hat{\mathcal{U}}_{h}(x)$ ($N$=5) without and with the $\mathrm{R_Z}(2\pi x)$ gate, respectively. Both cases show that Hartley states are orthonormal when $x = x'$, but the nonzero overlap residues in \textbf{(a)} are way too high, inhibiting model trainability. \textbf{(c)} $N$-qubit phase feature map and representative 2D squared overlap between normalized Fourier states $|\langle \tilde{\varphi}(x')|\tilde{\varphi}(x) \rangle|^2$ for $\hat{\mathcal{U}}_{\varphi}(x)$ ($N$=5). Notably, the Hartley overlaps are purely real, $\langle \tilde{h}(x')|\tilde{h}(x) \rangle \in \mathbb{R} $, unlike the complex overlaps between Fourier states, $\langle \tilde{\varphi}(x')|\tilde{\varphi}(x) \rangle \in \mathbb{C}$.
    }     
\label{fig:Overlap}
\end{figure}

Similar to Eq.~\eqref{eq:cas_properties}, the squared overlap is employed to examine the orthonormal behavior of different states of continuous variables. As illustrated in Fig.~\ref{fig:Overlap}(a), the squared overlap between mutual normalized Hartley states, $|\langle \tilde{h}(x')|\tilde{h}(x) \rangle|^2$, shows unity along the diagonal 45 degree line, meaning that the diagonal states are orthonormal as usual. However, those off-diagonal states, i.e. states along 135 degree line, reveal an unfavored behavior as they disrupt orthogonality in between nodes, contributing at least 50\% of the maximum value. This is quite different from the phase feature map \cite{Kyriienko2022} written in the Fourier basis (Fig.~\ref{fig:Overlap}c). This is mainly caused by the abrupt change in amplitude of the opposite states, making the model potentially difficult to train when using points between Hartley nodes. This can be fixed by introducing an additional $x$-dependent rotation that fixes overlap in between nodes. Specifically, by appending a $\mathrm{R}_\mathrm{Z}(2 \pi x)$ gate to the ancilla register (see Fig.~\ref{fig:Uh(x)}), we ensure a smooth transition between continuous states while retaining the real-valued amplitudes of $|\tilde{h}(x) \rangle$ over the entire domain with an off-diagonal contribution less than 5\% (Fig.~\ref{fig:Overlap}b). With $\mathrm{R}_\mathrm{Z}(2 \pi x)$ gate included, the resulting squared overlap looks similar to that in Fourier case.

\begin{figure*}[ht]
\begin{center}
\includegraphics[width=1.0\linewidth]{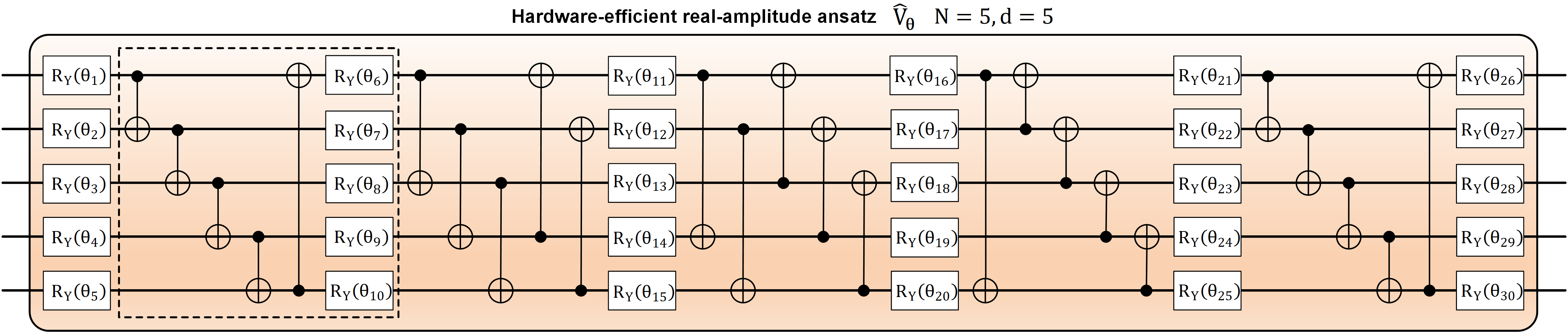}
\end{center}
    \caption{\textbf{Hardware-efficient real-amplitude ansatz (HERA) circuit.} HERA for \textit{N} = 5 qubits with the circuit depth of \textit{d} = 5. The first depth block $(d=1)$ is presented by a dashed box. The HERA is composed of $N(d+1)$ adjustable single-qubit $\mathrm{R}_\mathrm{Y}$ gates and $N \times d$ entangling (CNOT) gates. $\bm{\theta}$ is a set of training parameters. The entangling gates follow a modular arithmetic pattern that changes over the number of depth layers as follows: CNOT$[l, (l + m)$ mod $N]$ with $l \in [1,\cdots,N]$ and $m \in [1,\cdots,d]$.
    }
\label{fig:HERA}
\end{figure*}

\begin{figure}[ht]
\begin{center}
\includegraphics[width=1.0\linewidth]{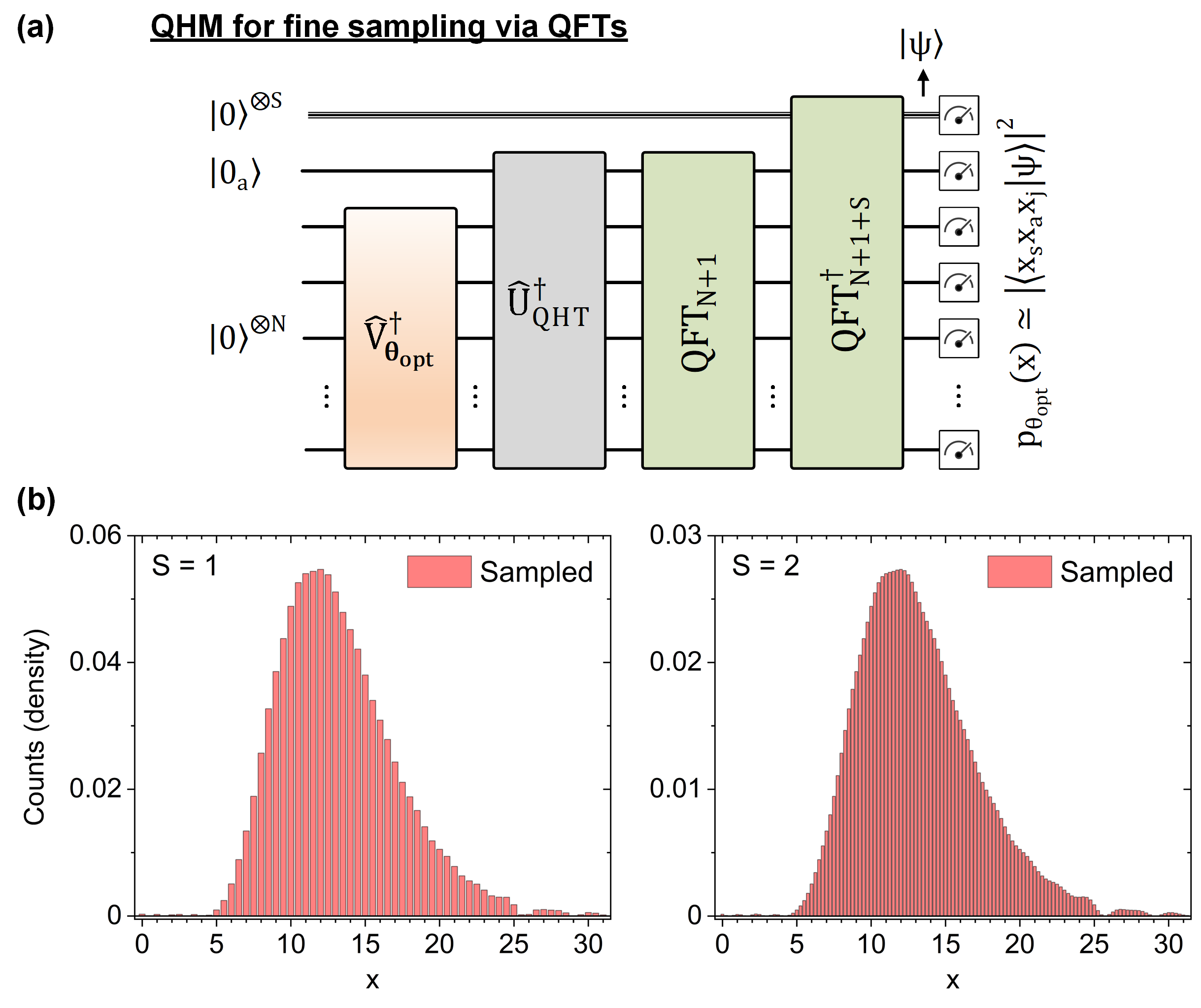}
\end{center}
    \caption{\textbf{Variant model for sampling in quantum Hartley-based models.} \textbf{(a)} Circuit with $S$-qubit extended registers on the top line used to perform dense sampling in the computational basis $|x_sx_ax_j\rangle$.
    Instead of a bitstring network, both QFT and extended inverse QFT circuits are used as components of the sampling circuit. Inverse QHT circuit is not extended with the same size as the regular ($S=0$) sampling case. \textbf{(b)} Sampled probability distributions from the same trained (GBM) model using the extended registers of $S=1$ and $S=2$ corresponding to double- and quadruple-frequency sampling, generated with $10^6$ and $10^7$ shots, respectively. The normalized histograms are plotted as a function of $x \in \mathbb{R}_{2^N-1}$.
    }     
\label{fig:SFig3}
\end{figure}

\begin{figure}[ht]
\begin{center}
\includegraphics[width=1.0\linewidth]{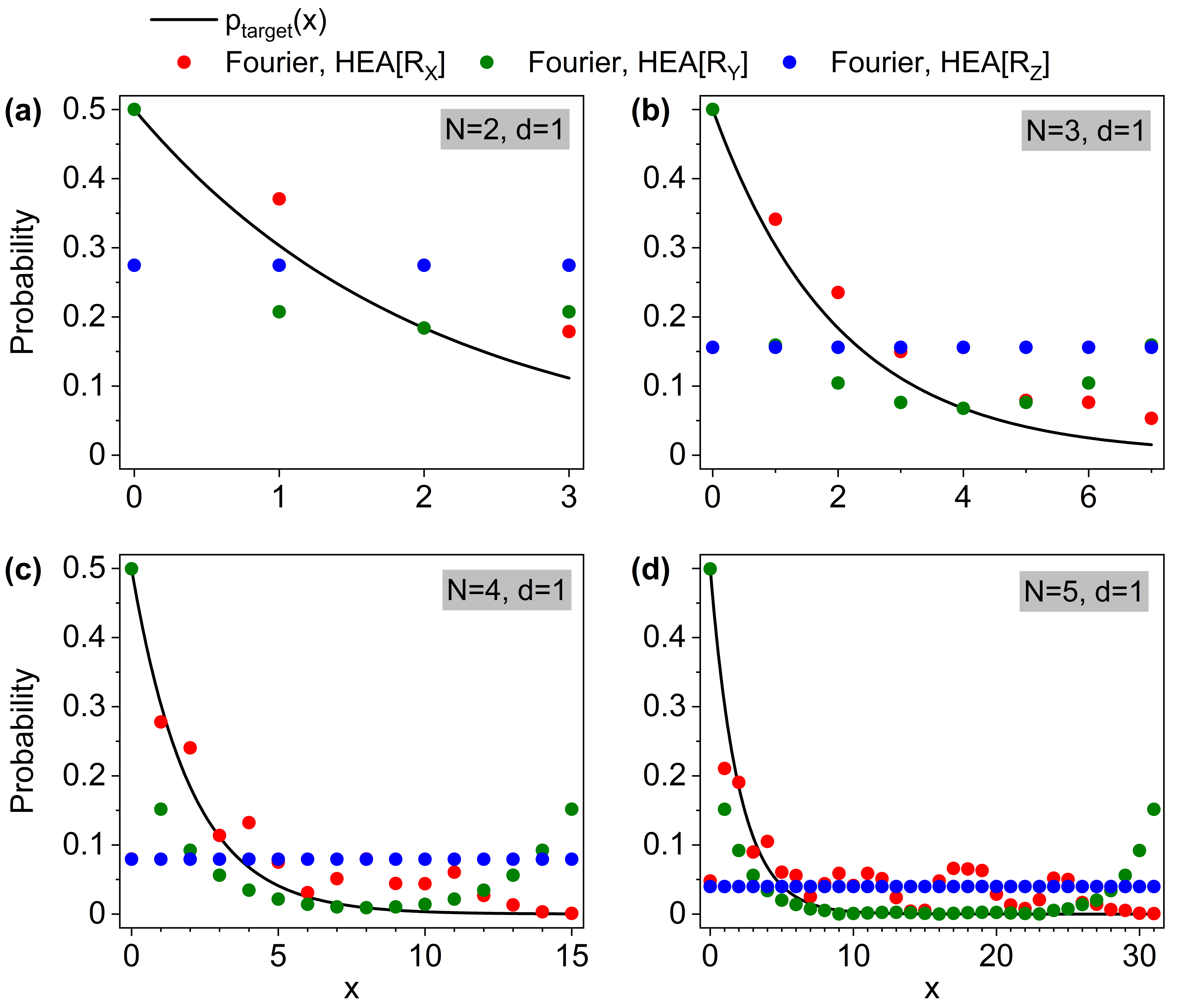}
\end{center}
    \caption{\textbf{Results of trained Fourier models.} \textbf{(a)--(d)} Training an exponential probability distribution with a scale inversely proportional to parameter $\lambda$ = 0.5, $ p_{\text{target}}(x) = \lambda e^{-\lambda x} $ for $x \geq 0$, using Fourier feature map circuits over $N$ = \textbf{(a)} 2, \textbf{(b)} 3, \textbf{(c)} 4 and \textbf{(d)} 5 qubits with the shallow HEA circuit depth of $d$ = 1. The parameterized element inside each depth block of the HEA is composed of one adjustable single-qubit rotation gate per qubit ($\mathrm{R}_\mathrm{X}$, $\mathrm{R}_\mathrm{Y}$, $\mathrm{R}_\mathrm{Z}$), and the entangling layers are the same as the HERA. These Fourier-based quantum models perform poorly due to an insufficient expressivity.  
    }     
\label{fig:SFig4}
\end{figure}

\begin{figure}
\begin{center}
\includegraphics[width=1.0\linewidth]{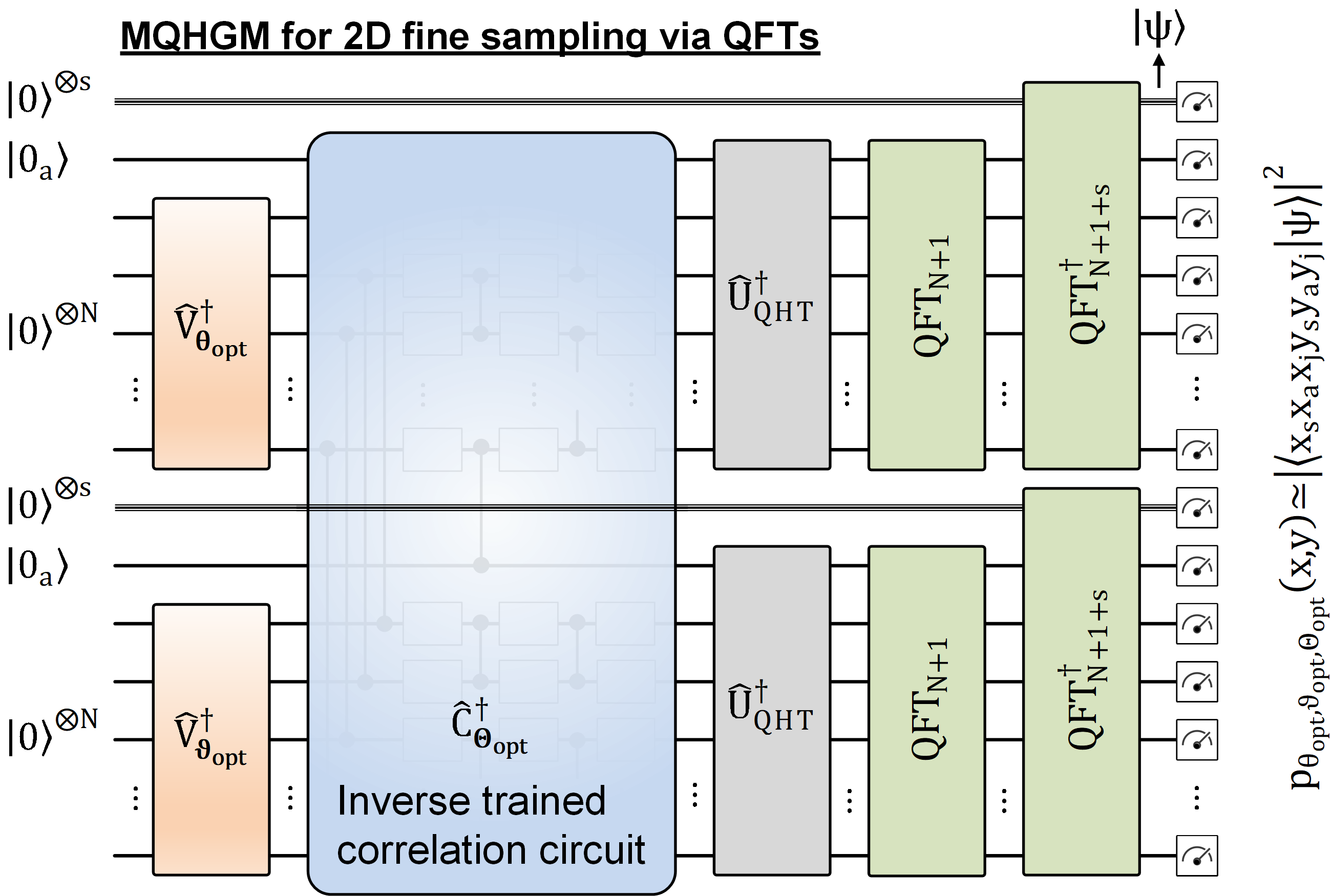}
\end{center}
    \caption{\textbf{Variant model for sampling in multivariate quantum Hartley-based generative models.} Circuit used to sample the multivariate distribution from the trained model, where $\bm{\theta _\text{opt}}$, $\bm{\vartheta _\text{opt}}$ and $\bm{\Theta _\text{opt}}$ are retrieved after the optimization procedure and two identical sets of inverse QHT, QFT and extended inverse QFT circuits are applied in parallel for dense sampling in the computational basis $|x_sx_ax_jy_sy_ay_j\rangle$. The quantum state just prior to measurement is denoted as $|\psi\rangle$.
    }     
\label{fig:SFig5}
\end{figure}

\end{document}